\documentclass[sigconf]{acmart}

\pdfoutput=1

\usepackage[noend]{algorithm2e}
\usepackage{amsfonts}
\usepackage{amsmath}
\usepackage{amssymb}
\usepackage{array}
\usepackage[english]{babel}
\usepackage{booktabs}
\usepackage{caption}
\usepackage{color}
\usepackage{comment}
\usepackage{enumerate}
\usepackage{epsfig}
\usepackage{etoolbox}
\usepackage{graphicx}
\usepackage{multirow}
\usepackage{subfig}
\usepackage{tabularx}
\usepackage{wrapfig}
\usepackage{xspace}

\ifdefined\isFinalized

\newcommand{\note}[1]{}
\newcommand{\mnote}[1]{}

\newcommand{\matt}[1]{}
\newcommand{\bobby}[1]{}
\newcommand{\peter}[1]{}
\newcommand{\riju}[1]{}
\newcommand{\abhishek}[1]{}

\newcommand{\textbefore}[1]{}
\newcommand{\textafter}[1]{#1}
\newcommand{\textalec}[1]{#1}

\else

\newcommand{\note}[1]{{\color{red}{\it #1}}}
\newcommand{\mnote}[1]{\marginpar{{\color{red}{\it\ #1 \ \  }}}}

\newcommand{\matt}[1]{{\color{blue}{\it Matt - #1}}}
\newcommand{\bobby}[1]{{\color{orange}{\it Bobby - #1}}}
\newcommand{\peter}[1]{{\color{purple}{\it Peter - #1}}}
\newcommand{\riju}[1]{{\color{green}{\it Riju - #1}}}
\newcommand{\abhishek}[1]{{\color{brown}{\it Abhishek - #1}}}

\newcommand{\textbefore}[1]{}
\newcommand{\textafter}[1]{#1}
\newcommand{\textalec}[1]{#1}

\fi

\newcolumntype{L}[1]{>{\raggedright\let\newline\\\arraybackslash\hspace{0pt}}p{#1}}
\newcolumntype{C}[1]{>{\centering\let\newline\\\arraybackslash\hspace{0pt}}p{#1}}
\newcolumntype{R}[1]{>{\raggedleft\let\newline\\\arraybackslash\hspace{0pt}}p{#1}}

\let\oldSim\sim
\renewcommand{\sim}{\raise.17ex\hbox{$\scriptstyle\oldSim$}}

\newcommand{\tblh}[1]{\textbf{#1}\xspace}

\newcommand{\sys}{SeCloak\xspace}
\newcommand{\secloak}{\sys}
\newcommand{\itwoc}{I$^2$C\xspace}
\newcommand{\sekernel}{s-kernel\xspace}

\newcommand{\nskernel}{NS-kernel\xspace}

\newcommand{\smc}{{\tt SMC}\xspace}
\newcommand{\TZ}{TrustZone\xspace}
\newcommand{\tz}{TrustZone\xspace}
\newcommand{\Android}{Android\xspace}
\newcommand{\android}{Android\xspace}
\newcommand{\airplane}{Airplane\xspace}
\newcommand{\movie}{Movie\xspace}
\newcommand{\stealth}{Stealth\xspace}
\newcommand{\ISTR}{{\tt str}\xspace}
\newcommand{\ILDR}{{\tt ldr}\xspace}

\ccsdesc[500]{Security and privacy~Mobile platform security}
\ccsdesc[300]{Security and privacy~Privacy protections}

\copyrightyear{2018} 
\acmYear{2018} 
\setcopyright{rightsretained} 
\acmConference[MobiSys '18]{The 16th Annual International Conference on
Mobile Systems, Applications, and Services}{June 10--15, 2018}{Munich,
Germany}
\acmBooktitle{MobiSys '18: The 16th Annual International Conference on
  Mobile Systems, Applications, and Services, June 10--15, 2018, Munich,
Germany}
\acmDOI{10.1145/3210240.3210334}
\acmISBN{978-1-4503-5720-3/18/06}

\begin{document}

\title{\sys: ARM Trustzone-based Mobile Peripheral Control}

\author{Matthew Lentz}
\affiliation{\institution{University of Maryland}}
\email{mlentz@cs.umd.edu}

\author{Rijurekha Sen}
\affiliation{\institution{Max Planck Institute for Software Systems}}
\email{rijurekha@mpi-sws.org}

\author{Peter Druschel}
\affiliation{\institution{Max Planck Institute for Software Systems}}
\email{druschel@mpi-sws.org}

\author{Bobby Bhattacharjee}
\affiliation{\institution{University of Maryland}}
\email{bobby@cs.umd.edu}

\renewcommand{\shortauthors}{Lentz et al.}

\begin{abstract}
Reliable on-off control of peripherals on smart devices is a key to security
and privacy in many scenarios. Journalists want to reliably turn off radios
to protect their sources during investigative reporting. Users wish to
ensure cameras and microphones are reliably off during private meetings. In
this paper, we present SeCloak, an ARM TrustZone-based solution that ensures
reliable on-off control of peripherals even when the platform software is
compromised. We design a secure kernel that co-exists with software running
on mobile devices (e.g., Android and Linux) without requiring any code
modifications. An Android prototype demonstrates that mobile peripherals
like radios, cameras, and microphones can be controlled reliably with a very
small trusted computing base and with minimal performance overhead.

\end{abstract}

\maketitle

\section{Introduction}

Personal smart devices are now central to communication, productivity,
health, and education habits of their owners. Whether they are carried
in a purse or pocket, or worn on a wrist, these devices form a hub for
location, activities, social encounters, and even health status. As
more people incorporate these devices in to their daily lives, more
personal data resides in these devices.

With the increasing reach of these devices has emerged new classes of
attacks on personal privacy in the form of data breaches from
malicious apps and OS compromises.  Data available to these devices
are often of a highly sensitive nature, including
biometrics~\cite{biometric1}, health
information~\cite{mhealth1,mhealth2}, user location and unique device
ID~\cite{androidleaks,pios,devident1}, user
activity~\cite{lovense,fitbitsex} including
conversations~\cite{iotspy1} and video~\cite{babyspying}, all of which
have been subject of attacks and leaks.
Beyond explicitly gathered data, the sensors on these devices can
often be configured or coaxed into providing sensitive information
such as user conversation, location, and activity beyond their
original design~\cite{powerspy,gyromic,accessory,pinskimming}.  The
fact that such attacks are now common is not surprising: as vendors
add more functionality, the software base that, often, must be trusted
has grown to millions of lines of code, and the data that can be
stolen is highly lucrative.  {As a result, there is a perceived loss
of trust in these devices: sophisticated (and increasingly regular)
users are no longer certain {\em exactly\/} what their device is
monitoring or what data is being gathered.}  Worse, there are many
situations when the loss of privacy translates to loss of security (or
worse), and these users need unambiguous and reliable methods to
control their devices.  Currently, the only fully reliable control is
to remove the battery if possible (or place the device in a Faraday
cage.)

At the same time however, these devices have highly sophisticated
hardware security built into their architecture, which is commonly used
to store biometric information and for financial transactions.  In
this paper, we address the following question:
\begin{quote} What is minimally required atop
 existing hardware primitives to give users secure and direct control
 over the sensors and radios in their devices, without affecting the
 functionality of the rest of the device, or changing the installed
 large software base?
\end{quote}
The system we present, \secloak, short for ``Secure Cloak'' would
allow users to, for example, verifiably\footnote{``Verifiably'' in our
context does not refer to a formal proof of correctness, but to the
fact that the secure software can (1) unambiguously notify users of
the state of the hardware, and (2) allow users to reliably control the
hardware state of their device.  We believe the small size \textalec{of} our TCB
makes it (more) amenable to formal analysis, but such verification is
out of scope for this paper.}
turn off all radios (e.g., WiFi, Bluetooth, cellular) and sensors (e.g., GPS, microphone),
using hardware mechanisms.  This guarantee would hold even if apps
were malicious, the framework (e.g., Android) was buggy, or the kernel
(e.g., Linux) was compromised.

An independent layer addresses an important part of the general
security and privacy problem by enabling users to reliably control the
availability of certain I/O devices, regardless of the state of vendor
or third-party supplied software (in our prototype: apps, Android, and
Linux).  It is easy to make the case for \secloak with regards to
privacy:
\secloak can be used to reliably turn off recording devices such as
cameras, microphones, and other sensors whenever users require such
privacy; users would continue to be able to use the rest of the
functionality of their device.  There are other situations where
\secloak is useful that is perhaps even more critical: for example, 
\secloak would enable a
journalist to reliably place their phone into a mode where their
physical whereabouts cannot be traced.  Given the importance and
ubiquity of mobile devices, we believe that a capability such as
\secloak, which enables secure and robust control over hardware, is not
simply an option but a necessity in many use cases.

\secloak is built using the ARM \TZ hardware security extension.  In order,  
our design goals are to:
\begin{enumerate}
\item Allow users to securely and verifiably control peripherals on their device
\item Maintain system usability and stability
\item Minimize the trusted code base
\item Not change existing software including the apps, frameworks or OS kernels
\end{enumerate}

As we will explain later, \TZ provides extensive support for running
individual CPU cores in ``secure'' and ``non-secure'' modes, and
allows for dynamic partitioning of the hardware into secure and
non-secure components.  An isolated trusted OS kernel can run in
secure mode, and control all memory/peripheral accesses and interrupts
received by the non-secure kernel.  The facilities provided by \TZ
makes satisfying (1) and (3) relatively trivial.  However, since the
non-secure software (in our case: Linux, Android, and
all apps) are not written for such dynamic system partitioning,
satisfying (1)-(4) simultaneously is \textafter{surprisingly} difficult.

The primary contribution of this paper is the design and
implementation of an end-to-end system that (nearly) satisfies all
four of our stated goals.  \textafter{\secloak employs a small custom
  kernel, called the \sekernel, to securely place the device in a
  user-approved state. The \sekernel traps and emulates load and store
  operations to protected devices by the untrusted Android platform to
  enforce the user's policy with a very small trusted computing base.}

Users can set preferences using an untrusted \Android app.  These
preferences are conveyed to the \sekernel by the app, and then
confirmed by the user to the \sekernel directly \textbefore{using a
  simple and novel technique}.  Beyond control of individual devices,
the app identifies different modes of operation (e.g., \airplane mode
that turns off all transmitters) for convenience. Even with parts of
the hardware not available to the non-secure kernel, \Android remains
usable, and the device performs as expected.  While the app itself is
rudimentary, we believe our proof-of-concept demonstrates that goals
(1) and (2) can be simultaneously satisfied.  The entire \sekernel,
including secure device initialization, setting device state per user
preference, user interaction for confirmation, and instruction
emulation, is only 15k lines of code. (In comparison, the Linux kernel
is roughly 13m lines of code.)  We believe our \sekernel TCB size
satisfies the third goal. Finally, \secloak requires 1 source line to
be changed in the Linux kernel (the change can be applied directly to
the kernel binary if the source is not available), introduces a new
kernel module (for calls into the \sekernel), and no change whatsoever
to other software layers, including \Android.  Thus, we get very close
to satisfying the last goal of not having to change existing software.

The rest of this paper is organized as follows. We cover related work and
background in ARM TrustZone in Section~\ref{sec:related} and describe the
design of the \sekernel in Section~\ref{sec:design}.  We describe our secure
kernel in Section~\ref{sec:sekernel} and non-secure software in
Section~\ref{sec:nskernel}.  We present an evaluation of \secloak in
Section~\ref{sec:eval}, and conclude in Section~\ref{sec:conc}.

\section{Background and Related Work}
\label{sec:related}

Motivated by the security and privacy problems in mobile devices,
recent work envisions many different solutions for malicious apps,
including novel permission
models~\cite{quire,appguard,hey-you,permission-analysis,apex-android-permission,curbing-android,taming-information-stealing}
and
sandboxing~\cite{boxify,android-sandbox,sandbox-comparison,javascript-iot}.
Reference monitors and security
kernels~\cite{vmmvax,seckernel,secopenflow,lsm,sechw,crepe,polkit,selinux}
provide fine-grained access-control mechanisms that can contain
application misbehavior.
However, these solutions assume that the kernel itself is not
compromised. To address problems with compromised or malicious
kernels, we have to consider approaches that use hardware-based
containment of peripherals.  Hence, in this section, we focus on
hardware techniques that can be used to implement an isolated software
component that controls peripherals.

Peripherals can be isolated from a platform OS using virtualization
extensions~\cite{overshadow,scc} or hardware security
extensions~\cite{intel-sgx,arm-tz,optee,scc,trustshadow}.  We discuss a
subset of such prior works, specifically focusing on secure peripheral
control.

\paragraph{Controlling peripherals via Virtualization}
One approach is to run the platform OS as a guest within a virtual
machine, leaving a hypervisor in control of peripheral devices.
Existing work uses hypervisors to isolate applications from untrusted
OSs~\cite{overshadow,inktag,sego,trustvisor}, protect integrity and
confidentiality of the application memory and provide support for
managing cryptographic keys.  Beyond memory integrity,
Inktag~\cite{inktag} and Sego~\cite{sego} provide trusted access to
the filesystem.  These systems are designed for isolating individual
applications from the OS, but don't provide a mechanism to reliably
control generic peripherals like \sys.

Zhou et al.~\cite{commodityx86, wimpy} propose trusted path schemes
using a hypervisor to host the untrusted OS and all trusted program
endpoints (PEs) of applications in separate VMs, with the PEs
supplying all of the necessary device drivers.
DriverGuard~\cite{driverguard} protects I/O data flows by using a
hypervisor to ensure that only privileged code blocks (PCBs) can
operate on the raw, unencrypted I/O data.  BitVisor~\cite{bitvisor} is
a hypervisor that implements ``para-virtualization'' for enforcing
security on I/O control and data.  SGXIO~\cite{sgxio} posits a system
in which a hypervisor hosts the untrusted OS (running in a VM) as well
as the trusted I/O drivers (running in Intel SGX~\cite{intel-sgx}
enclaves).  Lacuna~\cite{lacuna} ensures that I/O flows can be
securely erased from memory once they terminate, by relying on
virtualization, encryption, and direct NIC access.

These approaches seek to protect the integrity and confidentiality of
I/O data paths while maintaining full functionality, which comes at
the expense of a large TCB. \sys, on the other hand, provides reliable
on/off control of smart device peripherals based on a very small TCB.

\paragraph{Hardware Security Extensions}
Beyond virtualization, modern architectures offer trusted hardware
components e.g., Intel SGX~\cite{intel-sgx}
and ARM TrustZone~\cite{arm-tz}, which can be used to isolate software
components from an untrusted platform OS.  These techniques go beyond
Trusted Platform Modules (TPM), which enable secure boot, or Intel
Trusted eXecution Technology (TXT)~\cite{intel-txt} and AMD Secure
Virtual Machine (SVM)~\cite{amd-svm}, which allow for attested
execution of the OS or smaller code segments~\cite{flicker}.  Of
particular interest to mobile smart devices is TrustZone, because ARM
is the dominant CPU architecture in this market and TrustZone supports
the isolation of peripheral device access.

ARM TrustZone~\cite{arm-tz} is a set of hardware security extensions
that supports isolation of two ``worlds'' of execution: non-secure and
secure.  TrustZone differs from SGX in two important ways: TrustZone
does not address hardware memory attacks (i.e., it does not encrypt
the secure world's RAM), and unlike SGX, TrustZone addresses
peripheral device security as described next.

Each processor core executes in the context of a single world at any time; a
core can ``switch'' worlds using a privileged instruction (and, if
configured, upon exceptions or interrupts).
All accesses to memory and I/O devices are tagged with an additional bit,
the 'NS' bit, which specifies whether the access was issued while the core
was in non-secure mode.
Components in the system (e.g., bus and memory controllers) can be
configured, in hardware, to only allow secure accesses.

By default, TrustZone supports a single isolated execution environment
(the secure world), with a secure boot process that can be used to
verify the bootloader and the secure-world kernel.
Komodo~\cite{komodo} provably extends TrustZone to support attested
isolated execution environments similar to Intel SGX enclaves (though
physical memory snooping remains out of scope.)  Cho et al.~\cite{scc}
explore a hybrid approach to supporting isolated execution
environments with both a hypervisor and ARM TrustZone. During the
lifetime of a secure application, the hypervisor is active and
provides isolation; otherwise, the hypervisor is disabled (reducing
overhead) and TrustZone hardware protections are used to protect
sensitive memory regions.  Neither of these systems address peripheral
access control.

\paragraph{Controlling peripherals using TrustZone}
Prior work has built trusted path using TrustZone, with much focus on
balancing the size of the secure TCB versus functionality.  For
instance, TrustUI~\cite{trustui} enables trusted paths without
trusting device drivers by splitting drivers into an untrusted backend
and trusted frontend that runs within the secure kernel.  TrustUI uses
ad-hoc techniques, such as randomizing keys on the on-screen keyboard
after every touch, for ensuring that the information available to the
non-secure kernel does not leak device data.
ShrodinText~\cite{schrodintext} is a system for displaying text while
preserving its confidentiality from the untrusted OS.  ShrodinText
establishes a secure path between a remote (trusted) server and the
local framebuffer/display for this purpose, relying on TrustZone and a
hypervisor (for MMU and IOMMUs) to secure parts of the rendering
stack.
Liu et al.~\cite{trustedsensors} uses trusted device and bus drivers,
implemented in TrustZone and hypervisors, to attest and encrypt sensor
readings.  \sys instead focuses on reliable on/off control of peripherals on
mobile platforms with a very small TCB.

TrustZone is the basis for commercial
products~\cite{trustonic,knox,qsee,optee} that implement support for
isolated execution of secure applications and for secure IO.  Many of
these systems provide specific applications for secure data input
(e.g., PIN and fingerprint input, biometric
identification)~\cite{qsee, trustonic}.  Unlike \sys, these systems
provide no device configuration controls.  OP-TEE~\cite{optee} is a
small TCB OS for implementing secure applications over TrustZone.  Our
prototype \sys is built on a much pared-down version of OP-TEE.

\textafter{
Brasser et al.~\cite{securevenue} enable on/off control of peripheral
devices that are in restricted spaces (e.g., where the use of the camera is
not allowed).
The system relies on a local, TrustZone-isolated policy enforcement service
that grants a remote policy server read/write access to system memory.
To protect against rogue accesses, the user relies on a separate vetting
server that determines whether to allow or deny each of the policy server's
memory access requests.
The policy server uses this remote memory access capability to query the
state of the platform OS and modify the OS's device configuration according
to the desired policy.
The policy server must be able to handle any platform OS version,
configuration, and state, which increases its TCB and requisite maintenance
over time.
Also, the policy server cannot tolerate any vulnerability in the platform OS
that is not known to the policy server (e.g., zero-day exploits), and must
periodically re-check the state to ensure continued compliance.
Like this work, \sys provides reliable on/off control of peripheral devices;
however, it does so under a stronger threat model that includes a
compromised platform OS.
Additionally, \sys has a smaller TCB as it does not depend on any
details of the platform OS in order to meet the requisite security
properties.
}

\textbefore{
Brasser et al.~\cite{securevenue} provide on/off control of smart device
peripherals in restricted spaces where, for instance, the use of camera,
microphone, or radios is not allowed.  The system relies on
TrustZone-isolated policy enforcement code, which can authenticate itself to
a remote policy host server, and grant that server remote access to the
device's system memory. The policy server uses this remote memory access
capability to modify the untrusted OS's device configuration according to
the policy in effect, and periodically re-check the memory state for
modifications\footnote{A separate vetting service ensures the policy server
does not abuse its memory access to steal the user's confidential
information.}.  The server must have the ability to interpret and safely
modify the device's memory for any OS version, configuration, and state,
which increases its TCB and makes maintenance difficult in practice. The
system cannot tolerate zero-day exploits, or any vulnerability in the device
platform OS not known to the policy server.

\sys differs from this work in two ways. First, our aim is to provide the
device owner, not a third party, on/off control of peripherals. However,
\sys could be extended easily to delegate control, with the owner's consent,
to a third party. Second, \sys provides reliable on/off control under a
stronger threat model that includes a compromised platform OS, and with a
much smaller TCB, and without depending on details of a particular platform
OS version, configuration, and memory image.
}

\textafter{
Santos et al.~\cite{trustlease,trubi} present ``trust leases'', which allow
applications to request (with user approval) leases to place the device in a
restricted mode until some terminal condition is met (e.g., after 4 hours).
The trust lease model could be used to implement a settings application that
allows the users on/off control over peripheral I/O devices.
Their threat model assumes that the platform OS is trusted and correct;
their prototype implementation lives inside the Android framework and Linux
kernel.
In contrast, \sys has a stronger threat model that includes a malicious
platform OS, and operates as a separate, minimal secure kernel that runs
alongside the existing platform OS.
}

PROTC~\cite{protc} is a system for safeguarding flight control systems
on drones from non-essential but malicious software.  PROTC runs
applications in the \TZ non-secure world, and a kernel with access to
protected peripherals in secure word.  PROTC's secure kernel
communicates with ground control  using an encrypted channel.  Compared
to \secloak, PROTC is designed for a different application domain, and
does not allow for dynamic modification of protected peripherals.
Untrusted applications are always isolated, and the secure kernel
contains all of the protected device drivers.

\textafter{
Viola~\cite{viola} enables custom, per-peripheral notifications whenever the
I/O peripheral device is being used; for example, blinking the notification
LED when the camera is active.
Viola employs formal verification techniques to provide guarantees that the
user will be notified.
\sys and Viola are complementary: the user could use \sys to disable
devices and rely on Viola to notify them when specific enabled devices are
in use.
}

\newcommand{\smcSet}{{\small{\sc CLOAK\_SET}}\xspace}
\newcommand{\smcGet}{{\small{\sc CLOAK\_GET}}\xspace}
\newcommand{\ioctl}{{\tt ioctl}\xspace}
\newcommand{\wifi}{WiFi\xspace}
\newcommand{\bt}{Bluetooth\xspace}
\newcommand{\bitmap}{bitvector\xspace}
\newcommand{\dt}{device tree\xspace}

\section{Design Overview}
\label{sec:design}

\begin{figure}
\centering
\includegraphics[width=0.7\columnwidth]{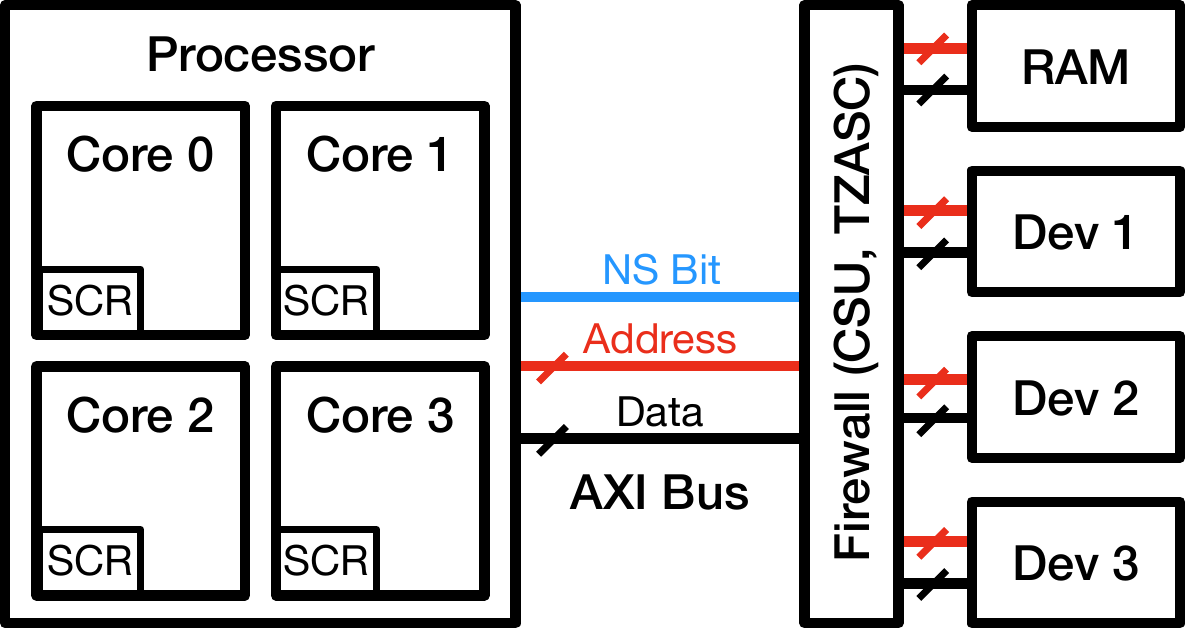}
\caption{\label{fig:arm-arch} High-level overview of SoC architecture,
  focusing on the security components that \secloak relies on.}
\end{figure}

\begin{figure*}
\centering
\includegraphics[width=\textwidth]{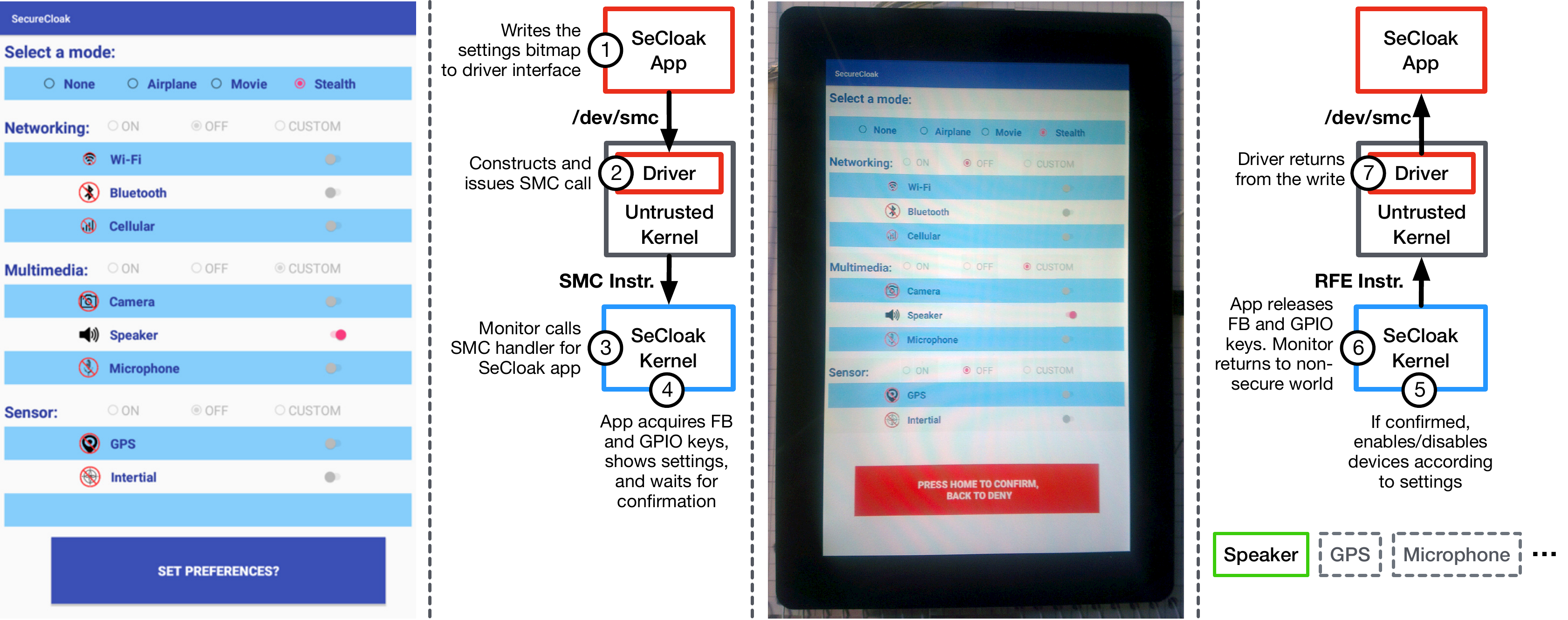}
\caption{\label{fig:workflow} Overview of the \secloak
workflow. The first panel (leftmost) shows a screenshot of \textalec{the} non-secure
Android app. The second panel shows the steps taken after the user
pushes the button to apply the settings. The third panel shows a
photograph of the secure \secloak app, which (re-)displays the settings to
the user and waits for user confirmation.  (Screenshots are not
possible in secure mode.)  The fourth panel shows the steps taken if
the user confirms the settings, such as disabling/enabling devices and
returning control back to the non-secure world (and app).}
\end{figure*}

As we have described, many prior approaches have ported (parts of) device
drivers into a secure kernel to provide robust access to peripherals, and
for limiting the access given to the \nskernel. However, none of them
satisfy the goals of \secloak.  Our work departs from these systems in two
major ways: we do not modify the \nskernel --- the \nskernel device drivers,
interfaces, etc. remain completely unchanged.  Next, our secure kernel TCB
is extremely small \textalec{(see Section~\ref{sec:tcbsize})}. Since the
\sekernel we describe here focuses entirely on providing the \secloak
functionality (as opposed to more general secure device access or control),
we do not require large device drivers or complicated mechanisms within the
\sekernel.  In fact, the bulk of the functionality provided by the \sekernel
relies on instruction trapping and emulation of ARM load and store
instructions to enable selective access to peripherals.

Figure~\ref{fig:arm-arch} shows a schematic of the ARM \TZ architecture (and
associated hardware components provided by the SoC we use to implement
\secloak).  Recall that the \TZ security model posits that the CPU can be in
a ``Non-Secure'' (NS) mode or ``Secure'' mode, and all memory and bus
accesses are tagged with the CPU mode using an extra ``NS-bit'' bus line.
(There are, in fact, many different privileged modes the processor can be
in, including a secure ``monitor'' mode that we use extensively.)  Different
peripherals, including parts of RAM, IO devices and interrupts, can be
configured to be available only in Secure mode.  In our design, Linux runs
as the \nskernel, and \secloak (with its secure kernel) controls the secure
modes of the CPU
\footnote{\textafter{Note that it is also possible to implement \secloak on
top of hardware virtualization mechanisms (i.e., nested page tables and
IOMMUs).}}
.

The ARM Security Configuration Register (SCR) contains the ``NS'' bit
that describes whether the core executing code is in the non-secure or
secure mode.  The SCR also contains a bit that determines whether a
hardware exceptions not raised by the CPU (so called ``external
aborts'' in ARM, e.g., a data abort raised from a memory access) cause
the processor to switch to the secure monitor mode.  As we shall see,
it is this setting of the SCR that enables \secloak to trap and
emulate instructions issued by the Linux kernel.
The SCR also contains similar configuration bits for interrupts, which
\secloak uses to listen for user input and support secure system reboot.
ARM supports two types of interrupts: IRQs and FIQs.
Traditionally, the secure world exclusively uses FIQs and the non-secure
world uses IRQs; therefore, the SCR is set to trap FIQs to the secure
monitor mode.

ARM provides a TrustZone Address Space Controller
(TZASC) that can partition portions of RAM such that they are available
only to secure mode accesses\textalec{, enabling isolation of the
\sekernel memory from Linux.}

Our \secloak prototype is built using a i.MX6 SoC.  i.MX provides a 
\TZ compatible component,  the  Central Security Unit (CSU), that extends  
the secure/non-secure access distinction to peripherals.  The CSU can
be used to enable secure-only access for different peripherals (which
\secloak uses), 
and also for programming access to various bus DMA masters (e.g., the
GPU).  The CSU contains a set of registers, called the ``Config
Security Level'' (CSL) registers, which can be set to mandate
secure-only access to various peripherals, e.g., GPIO controllers, PWM
devices, etc.  While our implementation programs the i.MX-specific
CSU, the design is general, and can be ported to other SoCs which provide
similar functionality.

In Figure~\ref{fig:arm-arch}, we encapsulate the TZASC and CSU as a
virtual ``firewall'' on the control path to RAM and devices.  In
hardware implementation, these components are not necessarily on the
control bus or the bus path, but may further program other hardware
components that control access.  For our purposes and for \secloak
software, this logical view of an intercepting firewall is sufficient
(and accurate).

\secloak uses a purpose-built kernel (the \sekernel) that runs in \TZ secure
mode. The \sekernel programs each of these components (the SCR, the TZASC,
the CSU) as required, both at initialization and runtime to enable \secloak.
We next discuss a user's typical workflow with \secloak before describing
how the \sekernel is configured to enable \secloak.

\subsection{Threat Model}

We assume that the device hardware is not malicious, and that the state
of the hardware (number and type of IO devices, their physical
addresses and buses, interrupts, etc.) is encapsulated in a ``device
tree'' file that is signed by a trusted source, as explained later in
Section~\ref{sec:devtree}.
(The hardware on modern embedded- and small-devices is usually
described in such a device tree, and existing kernels already provide
library routines to parse this format.)

Beyond the hardware, we assume that the boot ROM and bootloader is
trusted and correctly loads the \sekernel.  The \sekernel is also
trusted and assumed correct.

All other software in the system may be faulty or malicious.  This
includes any app the user may run, any framework layer (such as
\android), any kernel modules, and the \nskernel itself.

\subsection{\sys Workflow}

Figure~\ref{fig:workflow} shows the \secloak workflow.  Users interact
with a regular Android app (leftmost screenshot), and choose On or Off
settings for various IO devices and peripherals. This app is similar
to the ``Settings'' apps that are already available on smartphones and
tablets.

\textafter{An implicit assumption is that users understand which
  peripherals should be turned off (or on) under different
  circumstances.  Towards this end, the \secloak app helps users by
  providing pre-defined operating modes (e.g. \airplane mode or
  \stealth mode) and associated settings for the appropriate groups of
  peripherals.}  \textbefore{Of course, users have to trust the
  authors of the modes to have defined appropriate settings for the
  correct set of devices.}  As we shall see next, \secloak ensures
that for any mode or sets of individual devices that the user chooses
in this step, the user receives an unambiguous confirmation of their
state.

Once the user chooses ``Set Preferences?'', the app uses a Linux device driver to
invoke a cross-kernel call, called a \smc call in \TZ (second panel
from left in Figure~\ref{fig:workflow}).  The argument to this call
encodes the user's preferences.  The \sekernel receives these
preferences as part of the
\smc call handler.  Note that malicious software (the app, system services, the 
\nskernel) could have changed the preferences prior to the \smc call!  

User preferences, possibly modified, are received within the \sekernel
as part of the \smc call.  At this point, the \sekernel takes
exclusive access of the framebuffer and hardware buttons.  The
\sekernel parses the preferences, and recreates an image that exactly
corresponds to the settings chosen by the user, and copies this images
to the framebuffer. {\em If\/} the preferences had been modified, the
image on the framebuffer would {\em not\/} correspond to the settings
chosen by the user, and the user would notice a setting that does not
correspond to their choice.  The user is thus notified of malicious
software on their device.

The \sekernel changes the ``Set Preferences?'' button to one that allows the user
to confirm the settings by pressing a hard button (the ``Home''
button), or to go back (using the Back key) to the app and continue
changing settings.  We show a photograph of this screen (with a red
secure confirm button) in the third panel of
Figure~\ref{fig:workflow}.

A malicious app could display the preferences screen and then spoof
the confirmation screen.  It is imperative that during the
confirmation phase, the user is unambiguously notified that she is
interacting with the \sekernel.  Thus, during this phase, the
\sekernel  lights a protected LED which ensures the user that she
is interacting with the \sekernel.  This LED is never accessible to
the \nskernel.

Assuming the settings were as the user intended, she may confirm the
settings by pressing the ``Home'' button.  The \sekernel disables (or
enables) various IO devices and peripherals as instructed (rightmost
panel in Figure~\ref{fig:workflow}).

\section{\secloak Secure Kernel}
\label{sec:sekernel}

We describe \secloak's secure kernel, called the \sekernel, in this
section.  A signed device tree describes available hardware and
protections to the \sekernel.  Prior to describing the kernel itself,
we discuss how it is securely booted (next), and the device tree.

Modern devices are equipped with secure, tamper-proof, non-volatile
storage, into which device manufacturers embed (hashes of) public
keys.  The devices contain a one-time programmable Boot ROM that has
access to these keys, which are ``fused'' onto the hardware.

In bootstrapping \secloak, we assume that a trusted principal (either
the hardware manufacturer or the user) has performed the following steps:

\begin{itemize}
 \item Generated a trusted key, and stored the key onto the
 tamper-proof non-volatile storage.  For convenience, modern devices
 often allow multiple such keys to be ``fused'' onto the hardware;
 once installed, these cannot be removed or modified by software.

 \item Program the boot ROM to load \textalec{a signed
   bootloader image. The boot ROM verifies the signature against the fused
 key(s) and then executes the bootloader if successful.}

 \item \textalec{The bootloader ({\tt U-Boot}~\cite{uboot} in our case) contains a set of public
   keys which it uses to verify signatures on all loaded images. The
   bootloader will locate and load a signed \sekernel image and signed
   device tree blob (explained next). After verifying the signatures, the
  bootloader will execute the \sekernel. Modern bootloaders already support
  such verified booting.}
\end{itemize}

\subsection{Device Tree}
\label{sec:devtree}

\begin{figure}[t]
\centering
\includegraphics[width=\columnwidth]{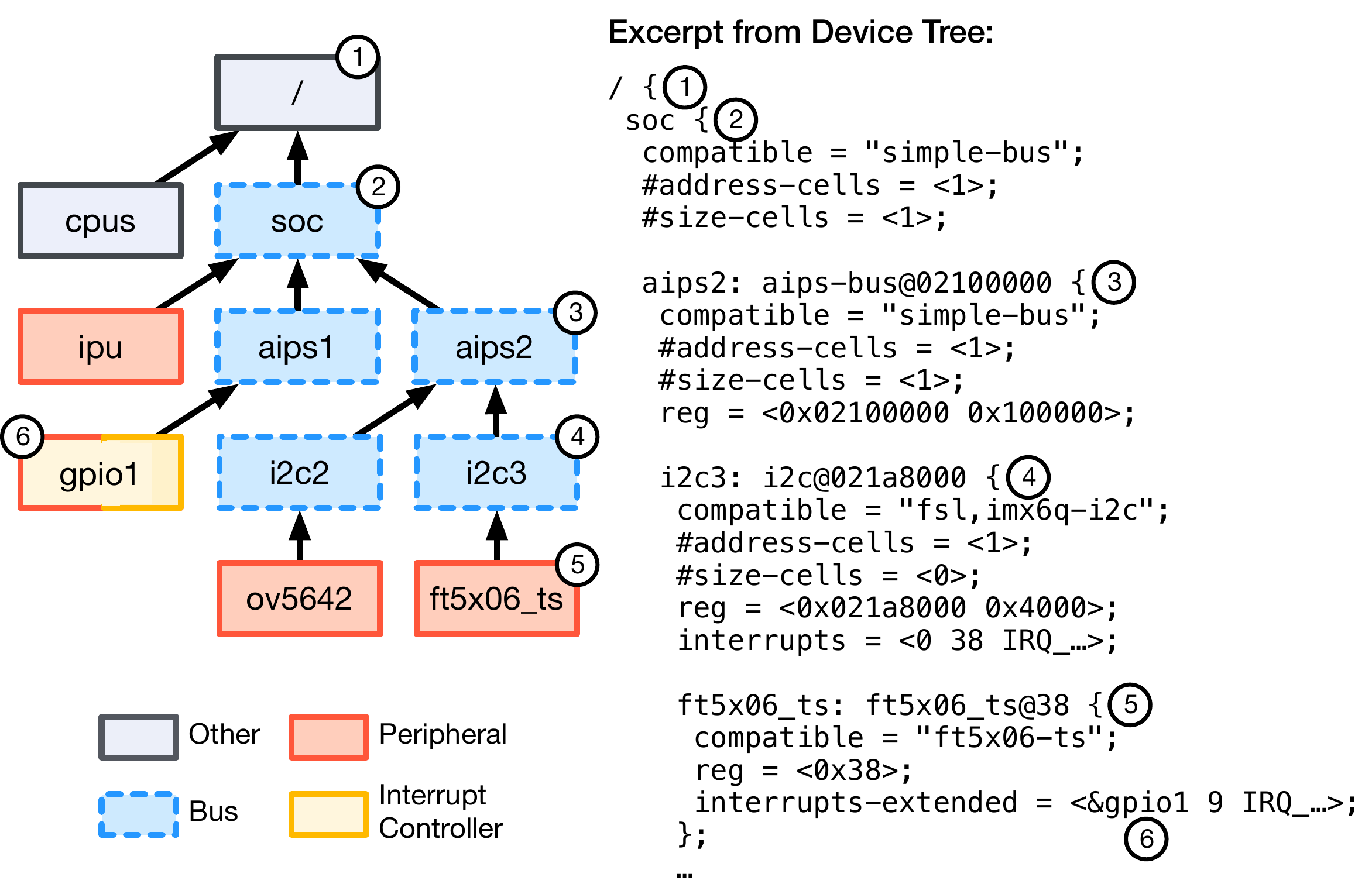}
\caption{\label{fig:devtree} Visualization of the device tree with a
  corresponding excerpt from the device tree file for the Boundary
  Devices Nitrogen6Q SoC. The labeled nodes correspond to parents and
  dependencies of the ``ft5x06\_ts'' touchscreen device.}
\end{figure}

The device tree structure~\cite{devtree} describes the hardware
devices present in a given system and how they are interconnected,
with each node representing an individual device.  It is important to
note that the device tree, by our assumptions, must be an accurate and
complete description of the hardware.  Otherwise, there may be other
peripherals in the system that can completely violate any and all
security properties, since the \sekernel would not know of their
existence, and would not be able to control the \nskernel's access to
these ``rogue'' peripherals.

Figure~\ref{fig:devtree} shows an excerpt from the device tree of our
prototype board, showing the arrangement of the touchscreen device
(ft5x06\_ts), the camera device (ov5642), the image processing unit (ipu),
and one of the gpio controllers (gpio1).  This a typical arrangement of buses
and devices on a modern SoC: for example, we see that the touchscreen device
and the camera are attached to \itwoc buses, which themselves are slaves of the
AIPS (AHB-to-IP-Bridge) which is a hardware bridge between the system bus
and third-party IP blocks such as the \itwoc controllers.

Each node has named properties along with a set of child nodes; some
properties might also express non-parental dependencies on other devices in
the system (e.g., touchscreen relying on GPIO pin for interrupts).
Some devices, such as buses, clocks, and interrupt controllers, have
required properties that denote how their children (for buses) or
dependents (for clocks and interrupt controllers) can reference their
resources.

In addition to the standard device tree, we add several properties to
device nodes.
First, we add a ``class'' property that maps low-level components
(such as interrupts and pins) to user-understandable names, such as
``microphone'', ``WiFi'', etc.  These class strings correspond to
individual devices that can be controlled via \secloak. 

Second, we add a ``protect'' property that identifies hardware
protection bits that must be set to protect the device.  On our
prototype, these map devices to their associated CSU registers.

\subsection{\sys Kernel}

Upon boot, the \sekernel initializes hardware defaults prior to
launching the \nskernel.  Specific steps include setting control and
security registers to appropriate defaults, and setting memory
protections such that the \nskernel cannot overwrite the \sekernel's
state.

The \sekernel initializes its internal data structures by initializing
the system MMU with virtual memory page table mappings for various
{regions}, including regions for non-secure RAM,
\sekernel  heap, and for  MMIO devices.
The \sekernel also starts the non-boot CPUs, and initializes per-core
threads and their contexts.  Faults and calls from the \nskernel
transition the CPU into a monitor mode, and the \sekernel initializes
the secure monitor with its stack pointer and call vector.  Finally,
the
\sekernel opens and parses the device tree.

\subsubsection{\sekernel Device Drivers}
\label{sec:drivers}

The \sekernel itself contains minimal drivers for three devices: GIC
(generic interrupt controller), GPIO controller, and the framebuffer.
The GIC driver isolates interrupts that can be received by the
\nskernel, the GPIO controller allows the \sekernel to directly 
interact with hardware buttons, and the framebuffer driver allows the
\sekernel to render the confirmation screen.    Together, they  enable the secure
part of \secloak.  We explain these drivers next.

\paragraph{GIC Driver}
The Generic Interrupt Controller (GIC) chip handles the distribution
of interrupts to CPU cores and enables isolated control and handling
of non-secure and secure interrupts. 
The \sekernel GIC driver supports functions to (1) enable or disable specific
interrupts, (2) set cpu mask and interrupt priority, (3) assign interrupts
to security groups, and (4) registering interrupt handlers. These functions
allow isolating interrupts associated with specific devices to be either
completely disabled or to be delivered to the \sekernel.  The \sekernel can
receive hardware interrupts and optionally re-deliver them to the \nskernel;
for example, this functionality is used by the GPIO and GPIO keys drivers
that we describe next.

\paragraph{GPIO Driver}
The general-purpose input-output (GPIO) controller supports input and output
operations on individual hardware pins.
In addition, the GPIO controller can also act as an interrupt controller on
a per-pin basis.
When an interrupt condition is triggered for a pin, the
GPIO controller triggers a (chained) interrupt which is handled by the GIC.

The \sekernel GPIO driver supports acquiring/releasing pins for exclusive
\sekernel  use, registering an interrupt handler for a given pin, and
reading (or writing) values from (or to) a pin.
The GPIO driver relies on the GIC driver to register its own interrupt
handler, which (when invoked) will read the GPIO device state in order
to determine which pins raised the interrupt, and then invoke handlers
corresponding to these pins.
The driver protects and emulates accesses to the GPIO controllers in order
to allow the non-secure world to continue to use any non-acquired pins while
preventing it from inferring any information about the acquired (secured)
pins.

Building on the GPIO driver, the GPIO keys driver supports hardware
buttons/keys connected to GPIO
pins \textafter{(e.g., power and volume buttons)}. 
The GPIO keys handler translates hardware button presses into a key
code that is specified in the device tree (e.g., \texttt{KEY\_BACK})
and passes it on to any \sekernel listeners.
The listeners can choose to consume the key press or allow it to be
passed back to the non-secure world.
We use the GPIO keys driver to register listeners for a secure
shutdown sequence (see Section~\ref{sec:reset}), and also for the cloak application to wait for the
user to confirm or deny the displayed settings.

\paragraph{Framebuffer Driver} 
The \sekernel framebuffer driver uses the image processing unit (IPU)
device to display  images.
When the \sekernel application acquires the framebuffer, the driver
allocates a single buffer in the secure region of memory and sets the
buffer format (RGB24) in the IPU.
Additionally, the driver protects access to the IPU and emulates accesses in
order to prevent the non-secure world from overwriting the settings (see
Section~\ref{sec:emudetail} for emulation policy details).
When the \sekernel application releases the framebuffer, the driver
restores the previous settings of the non-secure world and unprotects the
IPU.
Additionally, the framebuffer driver provides helper functions for clearing
the buffer with a single color and for blitting images onto the
display at specified locations.
We rely on the framebuffer driver for {(re-)}displaying the settings in the
\secloak app.  The images displayed by the \sekernel framebuffer driver
cannot be modified by the \nskernel.

\subsection{\smc Handlers}

The \sekernel supports two \smc calls, \smcSet and \smcGet, from the
\nskernel to enable \secloak.

The \nskernel invokes the \smcSet call with a \bitmap as the
argument.  Individual bits in the \bitmap correspond to the settings
for different device classes.  The \bitmap contains ``special'' bits
that encode modes (e.g., \airplane, \movie, \stealth), and groups
(e.g., Networking) as displayed by the app.  

The \smcSet handler executes the following steps:

\begin{itemize}
\item It starts by acquiring the framebuffer and GPIO keypad (via
GPIO).
As described in Section~\ref{sec:drivers}, acquiring devices applies
necessary hardware protection settings, emulation policy, and initial
settings for the secure use of the device.

\item The \smcSet handler parses the \bitmap and checks to see if it is valid;
if so, it uses framebuffer driver routines to blit corresponding images to
the screen in order to (re-)display the settings to the user.

\item Next, the notification LED, which is persistently acquired for exclusive use by
  the \sekernel, is turned on by the handler (via GPIO) to notify the user that the
\sekernel is in control.

\item The \smcSet handler then waits for the user to confirm (via the
`Home' button) or deny (via the `Back' button) the settings via its
registered GPIO keypad listener.
If the user confirms the settings, the handler will issue calls to enable or
disable each device class; otherwise, if the user denies the settings, the
handler does not take any action.
\item Finally, the handler releases the acquired devices (which resets per-device
state as necessary, e.g., framebuffer formatting and addresses) and
returns.
\end{itemize}

In order to disable (or re-enable) a device class, the \smcSet handler
first identifies all devices that belong to the given class (as
described in the device tree).
For each of those devices, the handler locates any ``protect''
properties, which identify the hardware protection that must be set to
isolate the device.  In some cases, the device itself may not have
hardware protection, but the bus it is located on may.  Thus, the code
must search for possible hardware isolation not just at the device
node, but recursively up the device tree as well.  In this way, the
\sekernel applies the hardware isolation for each device as described
in the device tree.

The \nskernel can use the \smcGet call to receive a \bitmap that
encodes the current protection state of device classes (and which mode
is active or which groups are enabled or disabled).  Upon launch, the
non-secure \android app uses this call to render an initial setting.

\subsection{Non-Secure and Secure Device Sharing}
We rely on the ability to share devices between the non-secure and secure
worlds, such as for providing a secure shutdown sequence via the GPIO keypad
\textafter{(e.g., power and volume buttons)} while still allowing the
non-secure world to handle button presses.
Two underlying mechanisms enable such sharing: 1) interrupt (re-)delivery to
the non-secure world, and 2) emulation policy to control the non-secure view
of the device.
While explaining the mechanisms, we will focus on the example of the GPIO
controller.
The GPIO controller uses a GIC interrupt in order to signal that the
interrupt condition is met for one (or more) pins.

In order to share interrupts with the non-secure world, we modify the device
tree such that the \sekernel operates on the actual hardware interrupt line
of the device that is connected to the GIC, while the \nskernel operates on
a (previously unused) interrupt line.
When the \sekernel receives \textbefore{the} an interrupt that should be
shared, it sets the corresponding non-secure world interrupt line pending
via the GIC.

\subsection{DMA}
Devices that are DMA masters can issue memory accesses on the system bus.
For example, the Image Processing Unit (IPU) will perform periodic DMA
transfers to read from framebuffers (whose addresses are specified in the
IPU's registers).
Each DMA master has permissions assigned for its bus accesses (i.e., non-secure or
secure), which (on our platform) are configured in the CSU registers.
In order to prevent the DMA masters from reading (or even modifying) the
\sekernel memory, we must configure their accesses as non-secure.
However, this presents a problem for the IPU device: since we need to
present a secure framebuffer to the screen, it must be able to perform
DMA accesses to secure memory regions.
To address this, we use the TZASC to configure the region that contains the
framebuffer as non-secure read and secure read/write.
While this lets the \nskernel inspect the secure framebuffer, we do not
require confidentiality of this framebuffer for any of our security goals
(only the integrity of its contents).

\subsection{Instruction Faults and Emulation\label{sec:emu}}

The \sekernel configures \TZ such that accesses by the \nskernel
to memory regions that belong to protected devices cause a fault.  This
fault is trapped by the monitor mode handler of the \sekernel.  We need these
traps to be able to selectively allow or deny \nskernel accesses to devices.

For a rudimentary
\secloak app, it is sufficient to simply configure \tz protections, 
and ignore these faults.  However, such a solution is unworkable if we
want the device to remain usable, as per our original goals, when
specific peripherals are protected.  In general, the \sekernel has to
trap the faulting instructions, and selectively emulate them based on
hardware state as we describe in this section.

There are two main reasons to intercept non-secure accesses to
protected resources and emulate these accesses in the secure world.
First, there can be a mismatch between the granularity of hardware
protection and that of individual devices that are being protected.
For instance, on the i.MX~6~\cite{imx6}, the Central Security Unit (CSU)
contains Config Security Level (CSL) registers that restrict access to
peripheral devices according to whether the accesses are made by the
non-secure or secure world.
These CSL registers group multiple devices into a single register
(e.g., {\tt GPIO1} \& {\tt GPIO2} or {\tt PWM1} through {\tt PWM4}).
If we want to disable a single (or subset of) devices, we must allow
accesses to all others that are protected by the CSL group.
Dependencies in the device tree can also cause mismatches in
hardware--software protection granularity.
For example, the ft5x06\_ts touchscreen uses a GPIO pin to signal an
interrupt to the processor when the user is touching the screen; in
order to secure the touchscreen, we must also secure the {\em
individual\/} GPIO pin, but not all the 64 pins that are protected by
the corresponding CSL register.

Second, we can use emulation for efficiently acquiring devices for
(temporary) exclusive use by secure applications, as well as to share
devices between the secure and non-secure worlds. This can reduce the
trusted codebase in the \sekernel, e.g., by allowing \nskernel writes
to the device for non-critical accesses.  We use this technique to
reduce the driver code size for the framebuffer driver.

\subsubsection{Instruction Emulation: Detail}
\label{sec:emudetail}
Each access to a device ultimately performs a memory-mapped
Input/Output (MMIO) read or write operation to a region of memory
associated with the device.  (The mapping of memory region to device
is obtained from the device tree.)
When hardware protections are enabled for a particular device, MMIO
accesses produce data abort exceptions; these are traditionally
handled by the \nskernel.

\paragraph{Hardware setup}
In order to intercept these accesses, \sekernel sets up the Secure
Configuration Register (SCR) in the CPU to specify that all external
aborts should be handled by the monitor.  This setting causes data
aborts to signal a fault that transitions the CPU into the secure
monitor mode.  The faulting address and related information is
available to the monitor fault handler.

In the \sekernel, the secure monitor fault handler invokes a routine
that determines whether to emulate or deny the access and, if
emulated, whether to modify the value being read or written.
The \sekernel maintains a data structure that contains regions of
memory (physical base address and size) corresponding to different
devices, along with the  prevailing policy for each.

The policy associated with each region may choose to deny a read or write. If a
read is allowed, the value that is read can be modified prior to being
returned to the \nskernel. If a write is allowed, the value to be written
can be modified prior to the write. The \sekernel code decodes the
instruction that cause the original fault: these instructions are of the
form \ILDR (load register for reads) or \ISTR (store register for
writes).  By default, the \sekernel emulates the instruction exactly and
returns control back to the \nskernel.

\begin{figure}
\centering
\includegraphics[width=0.9\columnwidth]{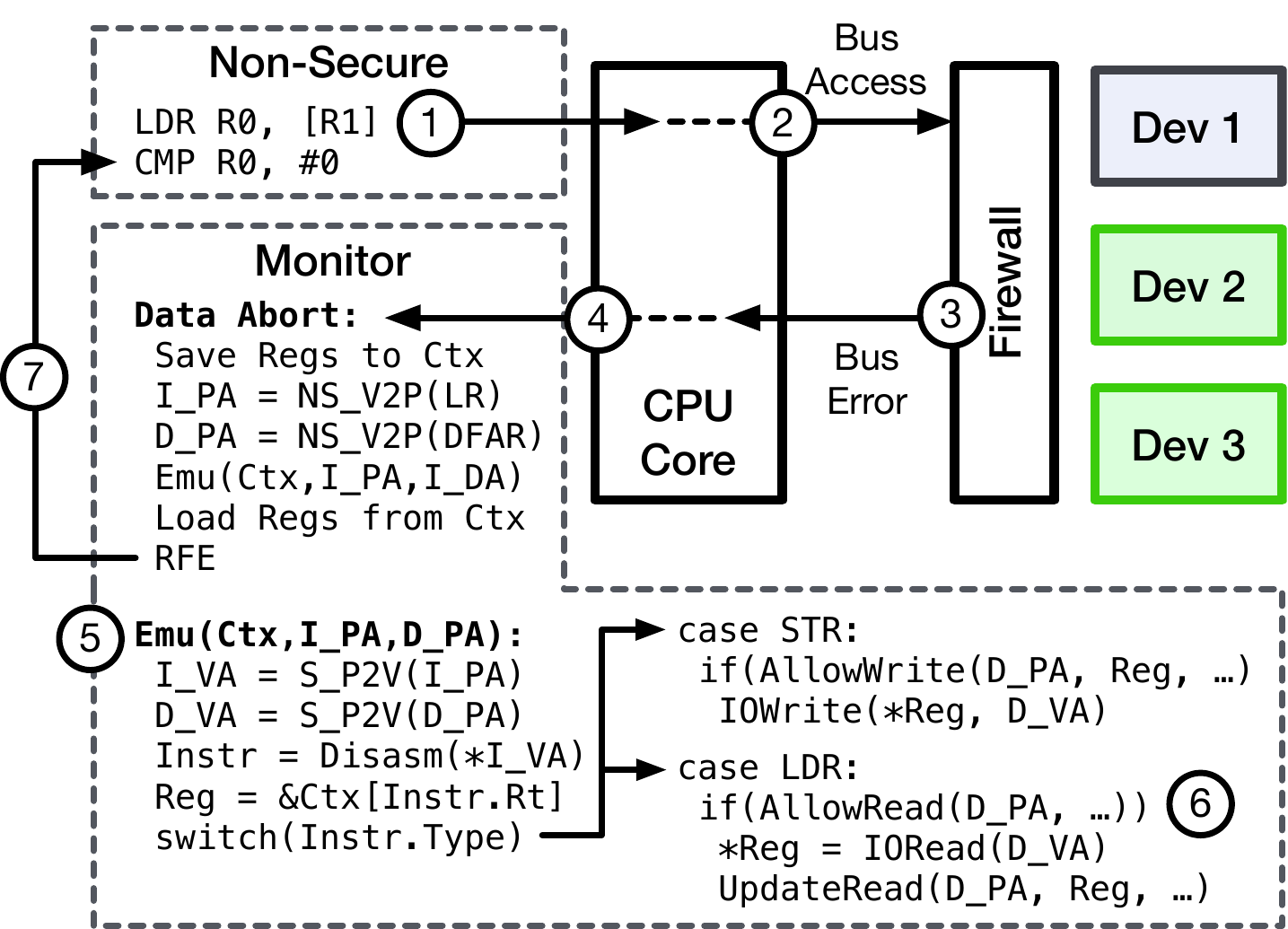}
\caption{\label{fig:emulation} Components and steps involved in intercepting
and emulating accesses made by the non-secure world.}
\end{figure}

\paragraph{NS-kernel Execution}
Figure~\ref{fig:emulation} shows this process.  Here a non-secure
driver attempts to read from a device (``Dev 1'') that is disabled by
\sys.  Specifically, the \nskernel code issues a \texttt{LDR R0, [R1]}
instruction to load from the address pointed to by \texttt{R1} into
\texttt{R0}.  The {\tt R1} register contains a memory address that
belongs to ``Dev 1''.

Upon executing this instruction (1), the CPU issues a read on the
system bus (2). This memory read is intercepted by the hardware
firewall (CSU/TZASC) responsible for protecting ``Dev 1''.
The firewall checks to see if the access should be allowed; if not, the
firewall returns a bus error to the CPU (which interprets this as an
external data abort).
In this case, since ``Dev 1'' is disabled by \sys, the firewall will
deny the read and report a bus error to the CPU (3).\footnote{The
system can be configured to also issue an interrupt to the CPU upon
such an error; we do not use this option in our
implementation.}
The CPU receives this bus error which corresponds to an external data abort.
Given the SCR configuration, the CPU switches to monitor mode and invokes
the monitor mode's data abort handler (4).

The data abort handler saves (and later restores) the current register set
\footnote{To be precise, \textalec{some registers in ARM are ``banked''
(e.g., the link register {\tt LR}), in that each mode has its own copy of the register}. The
abort handler saves the non-banked registers as well as the {\tt LR}
corresponding to the mode that caused the abort.}
and preserves the location to return to (i.e., the instruction
\emph{following} the faulting instruction).  

\paragraph{Fault Handling}
At this point, the fault handler has to determine two items: what was the
instruction that caused the fault, and what was the faulting address?  By
convention, the Data Fault Address Register (DFAR) contains the virtual
address of the access that caused the fault, and the LR register contains
the virtual address of the instruction that caused the fault. However,
these virtual addresses are \emph{non-secure} virtual addresses, and the
fault handler uses an ARM co-processor routine to resolve them into physical
addresses (``D\_PA'' from the DFAR, and ``I\_PA'' from the LR). The fault
handler then passes control to the \sekernel emulation routine with the
saved register context of the non-secure world (called ``Ctx'') and these
two addresses as arguments.

The emulation routine (5) begins by translating these physical addresses to
secure-world virtual addresses, and checks to make sure that they are in
appropriate regions: the non-secure RAM for the instruction and a device
MMIO region for the data address.
Next, the emulation routine invokes a custom instruction decoder we
have written to decode the instruction and determine the type of
instruction (whether it is a load or store) and the register
involved in the transfer.

Once the instruction and the physical address is decoded, the
prevailing policy (e.g., deny or allow with modifications) is
implemented as described above. In this case, the access is made by a
load instruction, so the emulation first checks to see if the policy
allows the read, performs the IO read operation, and finally checks to
see if the policy wants to modify the result (6). If allowed, the
final result is stored in the NS-context structure (that contains the
non-secure registers).

\textafter{In order to handle the case where multiple slave devices share a
  bus, a bus-specific policy must be provided. The device tree contains
  resource information that specifies how each device will be addressed on
  the bus; for instance, in the case of \itwoc, this corresponds to the
  7-bit slave address assigned to each device. The policy operates over the
  accesses to the bus's MMIO region to determine which device is being
  accessed, such that it can deny accesses to disabled devices (while
  allowing all others).}

\paragraph{NS-kernel resume}
The emulation routine then returns control back to the data abort handler,
which restores the registers for the non-secure world from the NS-context
data structure.  Note that for reads, one of these registers may now be
updated as a result.  Once the data abort handler terminates, the \nskernel
continues by executing the instruction directly after the one that caused
the data abort (7).

\section{Non-Secure Kernel}
\label{sec:nskernel}

Figure~\ref{fig:workflow} shows a screenshot of the \secloak \Android
app in the left-most panel.  The current version of the app is simple,
allowing users to set ON/OFF preferences for the devices on our
prototype board. Along with individual devices, the app allows users
to choose different operating modes (e.g., \airplane, \stealth) and
also provides the state of groups of peripherals (e.g., all networking
devices.)

Once the user presses the ``Set Preferences?'' button, the app invokes
a JNI call with a \bitmap that encodes the user
preferences.  The JNI module uses a Linux \ioctl call to pass the
\bitmap to the \secloak kernel module which, in turn, issues the
\smc call to the \sekernel with the \bitmap as an argument.

\subsection{A modification to the \nskernel}

Recall that our design goals were not to modify the \nskernel or
existing software if at all possible.  Unfortunately, without a single
byte modification, as we describe next, we can only provide the
security guarantee, but not maintain system stability.  

\secloak requires that the \sekernel be able
to trap individual accesses to protected devices and selectively
emulate these instructions.  However, as normally compiled, a Linux
binary on ARM does not raise data aborts that identify the {\em
specific\/} instruction that cause the abort.

Whether an instruction raises a precise or imprecise abort depends on
the page table entry (PTE) attributes of the memory that the
instruction attempts to access.
Precise data aborts are triggered for \ILDR and \ISTR instructions
that access ``strongly-ordered memory''. Strongly ordered memory does
not allow accesses to be buffered by either the processor or bus
interconnect~\cite{memorytype}.

As a result, we must modify Linux such that it configures its device
memory mappings to raise precise aborts.
While
we do this step directly in the source, it is a simple change that
can, in fact, be applied on the binary kernel image itself.

In our design, the \sekernel assumes that the \nskernel is
``compliant'' in setting device memory to be strongly-ordered.
However, a non-compliant \nskernel can still not access protected
devices.  It will, however, likely not receive any useful service from
protected device groups due to faulty emulation.

\paragraph{Kernel module}

The \secloak app requires a kernel module to invoke \smc calls, and we
have added such a module to Linux.  (Later versions of the Linux/ARM
kernels already provide a standard \smc interface like our kernel
module does, though even these kernels would require a module to
export a userspace interface.)  The kernel module provides a \ioctl
interface, which is used to communicate the user-selected \bitmap to
the \nskernel.

\paragraph{Framework Calls}

Along with the single change to the \nskernel, the \secloak app also
issues \android calls to address application and system stability.
Note that these are not {\em changes\/} to the framework, but instead,
extra calls that are invoked by the \secloak app.

When the user elects to disable certain devices, the \sekernel
configures hardware firewall mechanisms to prevent all accesses to the
disabled devices.
\textafter{
If the hardware device is attached to the system bus, then MMIO writes are
discarded and reads return 0; otherwise, if attached to a peripheral bus,
then bus access functions will return an error.
Ultimately, device drivers are responsible for handling these errors, which
typically involve several retries before abort.
These errors will further propogate to system services (and applications)
that are attempting to use the device; for example, when the camera is
disabled and the user attempts to run the camera app, an error message
appears after a few seconds.
}
\textbefore{Since neither the \android framework nor the Linux
kernel was designed to suddenly lose access to hardware, this step can
affect usability.}

Within the kernel, power management (PM) routines in device drivers
rely on the ability to communicate with their devices in order to save
relevant state and direct it to enter a low-power mode.
When a device is disabled by the \sekernel, these PM routines will
fail and thus keep the device in a high-power active mode.  In adverse
cases, the inability to transition individual devices into low power
states can prevent the entire system from being able to transition to
a low-power state, such as suspend-to-RAM.
Second, some device drivers may not contain appropriate error handling to
gracefully recover from errors resulting from the denied accesses.

Therefore, the \secloak app makes use of available system services
(e.g., WifiManager with {\tt setWifiEnabled}) to disable devices prior
to configuring the hardware firewall mechanisms (and likewise enable
devices after removing the hardware firewall restrictions).

\subsection{Device Reset}
\label{sec:reset}
\newcommand{\psci}{PSCI\xspace}

After a peripheral has been secured, malicious software inside the
\nskernel or framework can try to subvert security by rebooting the
entire device.  Such a reboot could happen without the user
necessarily noticing (while the device was idle) and could even be
remotely triggered.

One option is to make device policies persistent in the \sekernel,
such that they would be applied whenever the device is booted.  While
technically feasible (and indeed quite trivial), this option affects
usability.  Upon boot, the \nskernel (Linux) probes available devices
based on the device tree, and may not set up the device files and
other software correctly if the probe fails (which it would if the
device were secured upon boot.)  In turn, parts of the \android
framework may not initialize, leaving the device in a
unstable/unusable state.  Without a kernel rewrite or support, it is
difficult (if not impossible) to uniformly re-enable devices that were
protected at boot.

Instead, we adopt the following policy: the
\nskernel can reset the device only if there are no disabled devices.  
Otherwise, the \sekernel does not allow the \nskernel to invoke
\psci~\cite{psci} calls that are used to reset the processor.  

This design choice has the following implications: first, no code,
including remote exploits, in the \nskernel can reboot the device if
any peripheral is protected.  On the other hand, when the device is
rebooted, the regular \nskernel probes can proceed as usual, and the
device reboots in a fully usable state.  Further, the \sekernel does
not need to keep persistent state about policies, since the device
always reboots with all peripherals accessible to the \nskernel.  When
the \nskernel needs to reset the device (e.g., after OS or software
updates), the user must first run the \secloak app and remove all
protections.

The user may, at times, need to reboot the device after protection has
been applied. For instance, the \nskernel or \Android may become
unresponsive due to bugs or attacks.  To address this scenario, within
the \sekernel, we recognize a hardware key sequence that the user can
input to initiate a reset.  Since physical user input is necessary for
the device to be reset, this is a safe option, in that the user is
aware that the device is booting into a unprotected state.

\section{Evaluation}
\label{sec:eval}

We use the Boundary Devices Nitrogen6Q development board to run our
experiments, which contains an i.MX6 SoC with a quad-core ARM A9 processor
with TrustZone security extensions.
We use Android Nougat 7.1.1 with the Linux kernel version 4.1.15, both of
which are provided by Boundary Devices.
The \sekernel implementation is based on our custom fork of
OP-TEE~\cite{optee}.  OP-TEE is a OS for implementing secure
applications over TrustZone; \sekernel heavily modifies and reduces
the OP-TEE codebase.
Specifically, \sekernel retains OP-TEEs kernel threading and debugging
support.  \sekernel's MMU code is also based on OP-TEE.  The device
drivers required for \secloak (e.g., framebuffer and GPIO keypad),
device tree parsing, instruction interception and emulation, and the
code for securing device state was developed specifically for
the \sekernel.

We first present results to quantify the size of the TCB, both in terms of
the lines of code as well as the interface exposed to the \nskernel.
Next, we evaluate the overhead due to intercepting and emulating accesses
and show that, while there is a fair amount of overhead for individual
instructions, the reduction in overall system performance is negligible.

\subsection{Size of TCB}
\label{sec:tcbsize}

\begin{table}
\centering
\begin{tabular}{@{}lrrrr|r@{}}
& \multicolumn{4}{c}{\tblh{LOC Breakdown}} & \\
\tblh{Type} & \tblh{C Src} & \tblh{C Hdr} & \tblh{ASM} & \tblh{Total} & \tblh{Stmt} \\
\hline
Core & 3233 & 2357 & 1391 & 6981 & 3781 \\
&&&&\\
Drivers &&&&\\
\hspace{2mm} CSU & 45 & 9 & 0 & 54 & 29 \\
\hspace{2mm} Device Tree & 401 & 57 & 0 & 458 & 261 \\
\hspace{2mm} Frame Buffer & 146 & 29 & 0 & 175 & 113 \\
\hspace{2mm} GPIO & 562 & 15 & 0 & 577 & 284 \\
\hspace{2mm} GPIO Keypad & 169 & 14 & 0 & 183 & 89 \\
\hspace{2mm} $<$Other$>$ & 579 & 167 & 0 & 746 & 265 \\
\hline
Drivers Total & 1902 & 291 & 0 & 2193 & 1041 \\
&&&&\\
Libraries &&&&\\
\hspace{2mm} libfdt & 1220 & 350 & 0 & 1570 & 840 \\
\hspace{2mm} bget/malloc & 1421 & 68 & 0 & 1489 & 797 \\
\hspace{2mm} $<$Other$>$ & 1479 & 1182 & 81 & 2742 & 1212 \\
\hline
Libraries Total & 4120 & 1600 & 81 & 5801 & 2849 \\
&&&&\\
Total & 9255 & 4248 & 1472 & 14975 & 7671 \\
\end{tabular}

\caption{Breakdown of the lines of code (LOC) for different parts of our \sekernel
implementation. We list the LOC according to the language used (and
source vs. header) along with the total LOC. ``Stmt'' refers to
number of statements, which counts lines in assembly (ASM) and semi-colons
in C source and headers.}

\label{tbl:loc}
\end{table}

In Table~\ref{tbl:loc}, we show a breakdown of the lines of code for our
\sekernel implementation.
``Core'' consists of all non-driver and non-library code in the
\sekernel.
This code handles core \sekernel functionality, such as: memory management,
threading, the secure monitor, SMC handling (e.g., PSCI and CLOAK).
``Drivers'' consists of all driver code, which is further broken down into
specific drivers that we added to OP-TEE.
The ``$<$Other$>$'' category contains pre-existing drivers, such as the UART
(i.e., console), GIC, and TZASC-380 drivers.
The ``Frame Buffer'', ``GPIO'', and ``GPIO Keypad'' drivers are smaller than
their Linux counterparts since the secure drivers do not need to support all
device functionality.

As listed under ``Libraries'', our device tree parsing code relies on libfdt
to extract information from the flattened device tree file that the
bootloader places into RAM.
Additionally, the \sekernel uses the bget and malloc support for dynamic
memory allocation.
Finally, there are several other libraries and sets of functions that are
aggregated as ``$<$Other$>$'', such as: snprintk and trace functions (for
printing debug info), qsort (for sorting memory regions data structures),
and common standard library functions (e.g., memcpy, strcmp).
In general, we could further reduce our reliance on these libraries but
leave this for future work.

In total, our \sekernel comes to just under 15k LOC (\sim7.7k
statements).
The \sekernel has a limited attack surface in terms of the interfaces
that the \sekernel provides to the \nskernel, namely \smcSet and \smcGet.
\smcSet takes one argument, which is a bit vector containing the modes,
groups, and classes that the user wishes to enable or disable; \smcGet takes
no arguments.

\subsection{Emulation Overhead}

We perform two experiments to analyze the performance overhead introduced by
emulating non-secure instructions that access devices.
We focus on the case where the emulation is allowed, such as when devices
are shared between the non-secure and secure world (e.g., GPIO) or when
multiple devices (one of which is disabled) belong to the same hardware
firewall protection group.

\begin{table}
\centering
\begin{tabular}{@{}lcc@{}}
  & \multicolumn{2}{c}{\tblh{Instruction Time ($\mu$s)}} \\
  \tblh{Execution} & \tblh{Load (\ILDR)} & \tblh{Store (\ISTR)} \\
\hline
Linux & 0.11 & 0.29 \\
Linux+SOM & 0.27 & 0.33 \\
Emulated & 1.14 & 1.19 \\
\end{tabular}

\caption{Time to execute ARM instructions in the non-secure world that make
  device accesses. ``Linux'' execution uses the baseline Linux kernel
  without any changes. ``Linux+SOM'' execution uses the baseline Linux
  kernel but changing the device memory regions to enforce strong ordering
of accesses. For ``Emulated'' execution, we configure the \sekernel to
protect access to the WiFi controller and emulate the instructions that
result in data aborts.}

\label{tbl:micro}
\end{table}

In Table~\ref{tbl:micro}, we show the time taken to execute a single ARM
load (\ILDR) or store (\ISTR) instruction that access a 32-bit device register
on the WiFi controller.
We issued each instruction one million times to compute the time taken for
each individual instruction, and averaged this time over five trials.
We varied the execution between ``Linux'', ``Linux+SOM'', and ``Emulated''
modes.
For ``Linux'' execution, we use the baseline Linux kernel without any
changes, while for ``Linux+SOM'' we change the attributes for device memory
regions to enforce strong ordering of memory accesses (required for
interception and emulation, see Section~\ref{sec:emu}).
For ``Emulated'' execution, we configure the \sekernel to protect access to
the WiFi controller via the hardware firewall (i.e., CSU) registers and set
the emulation policy to allow accesses to the WiFi controller's MMIO
registers.

The requirement of strongly-ordered memory accesses imposes some overhead as
expected, increasing the time taken by 2.45x and 1.14x, respectively.
As expected, we see an increase in the time taken due to trapping and
emulating the instructions is 4.22x and 3.61x for loads and stores
respectively.
Note that, even though intercepting and emulating accesses incurs a fair
amount of overhead, high-throughput devices \textafter{(e.g., camera, network, and
display)} rely heavily on DMA transfers for performance and should remain
largely unaffected by emulation overhead (which only affects the control
path for DMA).
To that end, we next take a look at a macro-level benchmark involving the
WiFi controller.

\begin{figure}
\centering
\includegraphics[width=\columnwidth]{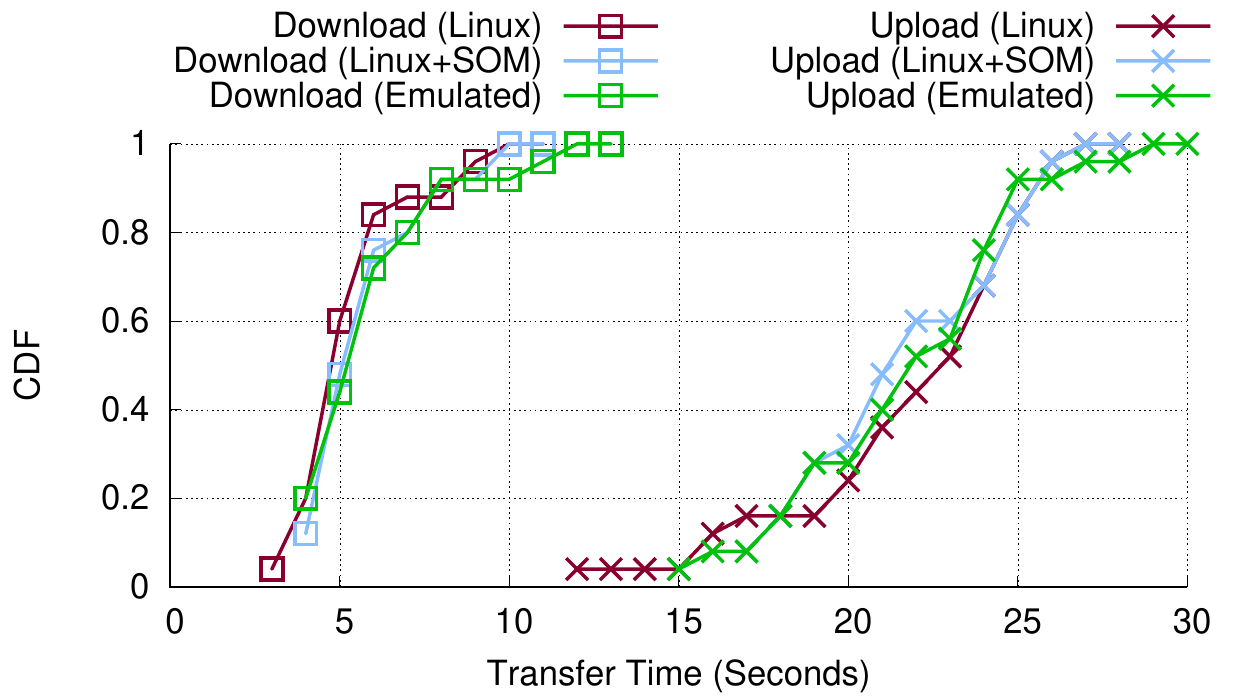}

\caption{\label{fig:time_cdf} Time taken for upload and download transfers of
a 10 MB file to complete over WiFi. ``Linux'', ``Linux+SOM'', and
''Emulated'' correspond execution modes evaluated in Table~\ref{tbl:micro}}

\end{figure}

Figure~\ref{fig:time_cdf} show the time taken to transfer files over WiFi
when the controller accesses are emulated vs. not.
We used the WiFi Speed Test~\cite{wifispeed} application to perform
the experiments and log the time taken for each of the trials; we used
a laptop as the other endpoint for the file transfers.  

The x-axis shows the time taken by the transfer in seconds, and the
y-axis shows the cumulative fraction of transfers that completed
within a given time.  Each CDF in Figure~\ref{fig:time_cdf} is
computed over 25 runs.

The download and upload performance shows that there is no visible
impact of interception and emulation on WiFi transfers, despite an
appreciable increase in execution time for individual load and store
instructions (as shown in Table~\ref{tbl:micro}).  This is because the
WiFi driver and controller, like all modern bulk data transfer
devices, uses DMA to transfer packets.  Once the controller firmware
is loaded, and the DMA tables configured, each packet transfer (which
can be many thousand bytes) requires very few (tens) MMIO instructions
to initiate the DMA.  We believe this result indicates that \secloak
can be used, even for high performance peripherals, without
significant impact on user-perceived performance.

\section{Conclusion}
\label{sec:conc}

In this paper, we have described a system, \secloak, that uses a
small-TCB kernel to allow users to unambiguously and verifiably
control peripherals on their mobile devices. Such a capability has
many uses, e.g., it can allow users to ensure they are not being
recorded, or journalists to ensure that they are not being tracked by
using radio or other means.

The main technical challenge in designing \secloak was to ensure that
existing mobile device software, in particular \Android and Linux,
could co-exist with the secure kernel without code modification and
without affecting device stability and usability.  Towards this end,
we have described an instruction emulation mechanism that enables
\secloak without changing existing software using a very small secure kernel.

\secloak is a system that allows users to assert binary control over
peripheral availability.  It is easy to imagine situations where finer
grained control is more appropriate, e.g., controlling the GPS device to
provide city-level location to specific apps, and true location to others.
In future work, we will extend our architecture to support non-binary
control over peripheral devices.
\textafter{In addition, we plan to explore more cooperation between the two
kernels for enabling device power management operations and for improving
performance by requiring strongly-ordered accesses only for protected
devices.}

The source code for our implementation is publicly available at:
\begin{equation*}
\texttt{http://www.cs.umd.edu/projects/secureio}
\end{equation*}

\section*{Acknowledgements}
We would like to thank our shepherd, Alec Wolman, and the anonymous
reviewers for their valuable feedback. The work was supported in part
by the European Research Council (ERC Synergy imPACT 610150), the
German Science Foundation (DFG CRC 1223), and the National Science
Foundation (NSF CNS 1526635 and NSF CNS 1314857).


\end{document}